\documentclass[prl,showpacs,superscriptaddress,twocolumn]{revtex4-1}

\usepackage{graphicx}
\usepackage{amssymb}
\usepackage{hyperref}\hypersetup{colorlinks=true,linkcolor=blue,citecolor=blue,pdfstartview=FitH}
\usepackage{bm}\renewcommand{\mathbf}{\bm}

\newcommand{\figwidth}{0.95\columnwidth}

\begin{document}
\title{Critical behavior of the Ashkin-Teller model with a
line defect: a Montecarlo study}
\author{G.~Duchowney}
%
\author{C.~Na\'on}
\author{A.~Iucci}
\address{Instituto de F\'{\i}sica La Plata (IFLP) - CONICET and Departamento de F\'{\i}sica,\\
Facultad de Ciencias Exactas - Universidad Nacional de La Plata, CC 67, 1900 La Plata, Argentina}

\begin{abstract}
We study magnetic critical behavior in the Ashkin-Teller model with an asymmetric defect line. This system is represented by two Ising lattices of spins
$\sigma$ and $\tau$ interacting through a four-spin coupling $\epsilon$. In addition, the couplings between $\sigma$-spins are modified along a
particular line, whereas couplings between $\tau$-spins are kept unaltered. This problem has been previously considered by means of analytical
field-theoretical methods and by numerical techniques, with contradictory results. For $\epsilon > 0$ field-theoretical calculations give a magnetic
critical exponent corresponding to $\sigma$-spins which depends on the defect strength only (it is independent of $\epsilon$), while $\tau$-spins
magnetization decay with the universal Ising value $1/8$. On the contrary, numerical computations based on density matrix renormalization (DMRG) give,
for $\epsilon > 0$ similar scaling behaviors for $\sigma$ and $\tau$ spins, which depend on both $\epsilon$ and defect intensity. In this paper we
revisit the problem by performing a direct Montecarlo simulation. Our results are in well agreement with DMRG computations. We also discuss some
possible sources for the disagreement between numerical and analytical results.

\end{abstract}

\maketitle

\section{Introduction}

Despite being massively studied for decades, two dimensional lattice spin models keep attracting great interest as a toolbox for the understanding of
phase transitions and critical phenomena. In particular, when defects are present and systems lose translational invariance, critical behavior becomes
nontrivial and physical properties on the defects can be different from those in the bulk ~\cite{Igloi_etal}. In addition, the critical properties of
these models are significant not only from an academic point of view, but are also relevant to fields as diverse as
biology~\cite{chang08_biology_applications} and the physics of cuprates in condensed matter systems~\cite{aji0_varma_cuprates_applications}.

Some of these models, such as the Ashkin-Teller (AT)~\cite{ashkin43_teller_model} and the eight-vertex model~\cite{baxter71_8V_model} have a very rich
phase diagram, which features partially ordered intermediate phases, various first-order and continuous phase
transitions, and exhibit as a salient feature non universal critical behavior, i.e., the critical exponents of certain operators are continuous functions
of the parameters of the Hamiltonian. More recently, many studies have focused on the quantum version of the AT model as a prototypical model for the
analysis of the efficacy of various sophisticated renormalization schemes~\cite{obrien15_AT_renormalization,bridgeman15_quantum_AT_MERA}.

A particularly interesting and fertile arena is the study of the role played by defects on local characteristics of these type of models, such as the
local magnetization and the correlation functions, though much less is known in these inhomogeneous cases. In the paradigmatic Ising lattice with a line
defect the critical exponent of the magnetization depends continually on the defect
strength~~\cite{bariev79_ising_line_deffect,mccoy_80_perk_continuous_exponents_ising}, whereas the
scaling index of the energy density at the defect line remains unchanged.

This problem was considered in a more complex system such as the AT lattice with a line defect, in ~\cite{AT_naon}. Let us recall that the AT lattice can
be viewed as two Ising lattices with spin variables $\sigma$ and $\tau$, respectively, interacting through their corresponding energy densities. Thus, in
the absence of this interaction one has two independent Ising systems. In \cite{AT_naon} an asymmetric line defect, affecting only one type of spins (to
be definite let us say the $\sigma$ spins) was introduced, and the critical behavior of spin-spin correlations was determined, through field theoretical
methods, for both $\sigma$ and $\tau$ spins. The magnetic critical exponents where found to be independent of the coupling between the Ising models. More
specifically, $\sigma-\sigma$ correlations decay as in Bariev's model \cite{bariev79_ising_line_deffect} whereas $\tau-\tau$ correlations behave as in the
usual (homogeneous) Ising model, with the universal $1/4$ exponent. These results stimulated a numerical study of the local critical behavior at an
asymmetric defect in the AT model \cite{AT_lajko}. By using density matrix renormalization, in the region of parameters where the numerical computation
can be compared with the field-theoretical results, these authors found that magnetization exponents at both $\sigma$ and $\tau$ spins are both dependent
on the interaction between Ising spins and the defect strength. These conclusions are in clear contradiction with the field-theoretical calculations of
\cite{AT_naon}.

The discrepancies described above call for re-visiting the critical properties of the the AT model with a defective line. Here we report on the numerical
measure of the AT model with a line defect over self-dual critical line with nonuniversal exponents that separates the ferromagnetically ordered phase
from the completely disordered one. We make use of the simple Metropolis algorithm to compute critical exponents. Our results are in well agreement with the DMRG study of \cite{AT_lajko}.

\section{The model}
The AT model can be represented as two overimposed copies
of the Ising squared lattice coupled by means of a four spin interaction
\begin{equation}
H_{\mathbf{AT}}=-\sum_{\langle\mathbf{r}\mathbf{r}'\rangle}J[\sigma_{\mathbf{r}}
\sigma_{\mathbf{r}'} +\tau_{ \mathbf { r }}
\tau_{\mathbf{r}'}]+J_4\sigma_{\mathbf{r}}\sigma_{\mathbf{r}'}\tau_{\mathbf{ r
}} \tau_{\mathbf{r}'}
\end{equation}
where $\mathbf{r}=(i,j)$ labels the lattice sites, $\sigma_{\mathbf{r}}$ and
$\tau_{\mathbf{r}}$ represent both Ising
spins, $\langle\ldots\rangle$ indicates a sum over nearest neighbors and $J_4$
represents the coupling.

We introduce an asymmetric line defect located at $j=0$ by modifying the coupling between $\sigma$ spins over a single line,
\begin{equation}
H_{\mathrm{defect}}=
-J_l\sum_{\langle\mathbf{r}\mathbf{r}'\rangle}\delta_{i,0}\sigma_{\mathbf{r}}
\sigma_{\mathbf{r}'},
\end{equation}
where now the effective coupling over the defective line is given by $J+J_l$.
Periodic boundary conditions are assumed in both directions. As it is known,the phase diagram
of the AT model is very rich. We shall be interested in the critical line
defined by the equation
\begin{equation}
 e^{-2K_4}=\sinh2K
\end{equation}
where $K=J/k_BT$ and $K_4=J_4/k_B T$, being $T$ the temperature, which
separates the ferromagnetic and paramagnetic phases for $K<K_4$, (for $K>K_4$
an intermediate, partially ordered phase appears). Over this line the clean system
exhibits non-universal critical behavior. We performed all the calculations
over this critical line.

We shall consider the critical behavior of the spin correlations on the
defect line
\begin{eqnarray}
 \langle\sigma_{(0,j)}\sigma_{(0,j+r)}\rangle\sim r^{-2x^\sigma},\\
 \langle\tau_{(0,j)}\tau_{(0,j+r)}\rangle\sim r^{-2x^\tau}.
\end{eqnarray}
In absence of the line defect, these exponents take the universal values $x^\sigma=x^\tau=1/8$ \cite{kadanoff77_critical_exponents_AT,
kadanoff79_correlation_functions_critical_AT}. Another limit that deserves attention, since it will be useful as a checking, corresponds to the case in
which the defect line is present but the Ising lattices are decoupled, i.e. $\epsilon=J_4/J = 0$. In
this case, the system reduces to two decoupled Ising models, one of them (the one identified with $\sigma$ spins) defective. The Ising model with a
defective line was studied by Bariev~\cite{bariev79_ising_line_deffect} and he found that
\begin{equation}
x^{\sigma} = \frac{2}{\pi^2}\arctan^2 e^{-2K_l},\label{eq:bariev}
\end{equation}
where $K_l=J_l/k_BT$. 
The $\tau$ spins become independent, conform a clean Ising plane and therefore
$x^{\tau}=1/8$.

\section{Numerical results}

We performed Montecarlo simulations on the square lattice of size $L\times L$ with periodic boundary  conditions and considered values up to $L=128$. We
analyzed several values of the defect intensity and the coupling between Ising planes. Instead of working with correlation functions, we computed the size dependence of the magnetization for the
$\sigma$ and $\tau$ spins. According to finite size scaling theory these magnetizations behave asymptotically as
\begin{equation}
	\label{eq:magvsL}
	m_{l}^{\alpha}(L)\sim L^{-x_{m}^{\alpha}}, \; \; \; \; \alpha=\sigma ,
\tau .
\end{equation}
For small system sizes, the behavior of the magnetization departs from the power-law as can be appreciated in Figs. \ref{fig:ml_vs_L_all} and
\ref{fig:ml_vs_L_neg_all}. However, for larger systems, typically for $L>64$, the power-law is restored and fits of the size-dependent magnetization in
those regions allow the extraction of the exponents.

\subsection{Decoupled ($\epsilon=0$) case}

\begin{figure}[h!]
\begin{center}
\includegraphics[width=\figwidth]{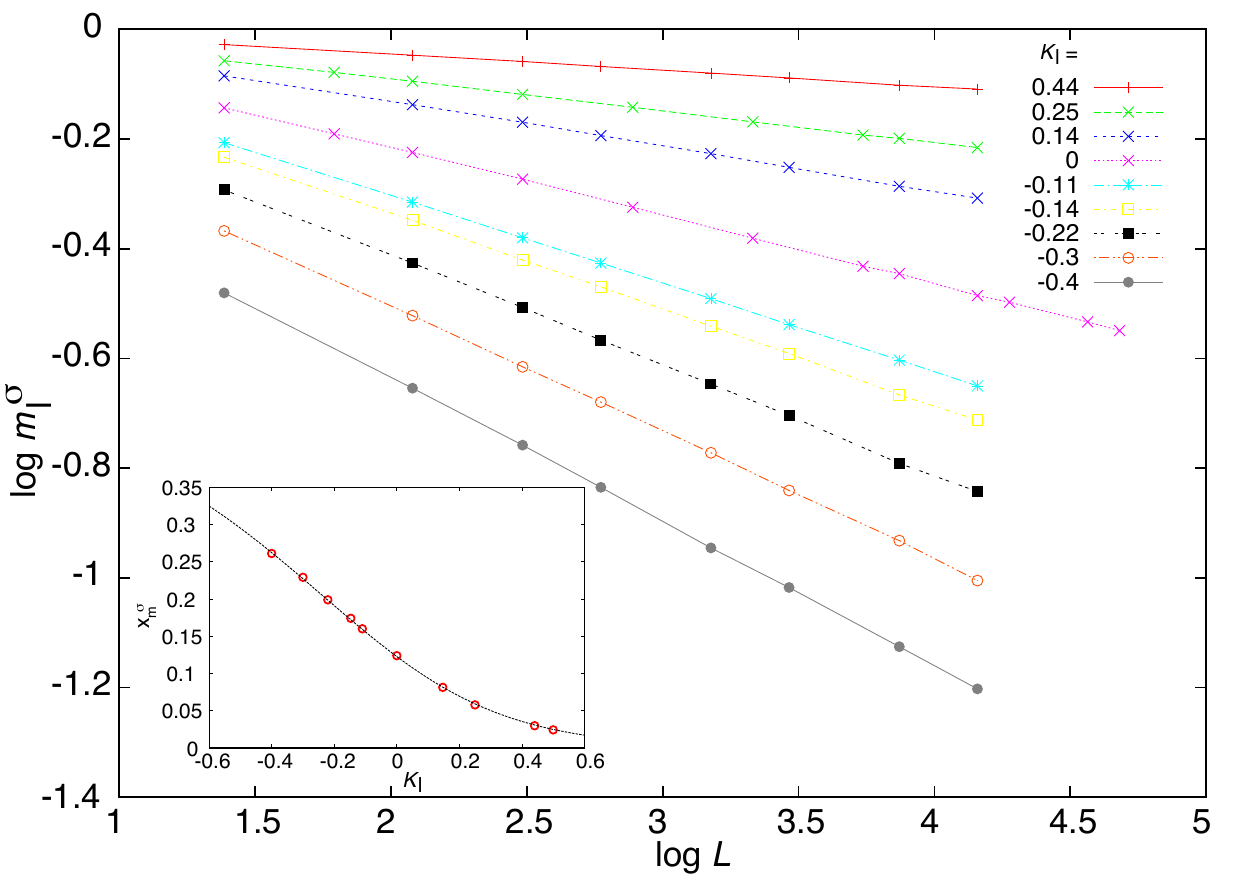}
\end{center}
\caption{Logarithm of the magnetization over the defect line as a function of the logarithm of $L$ for different values of $K_{l}$. Inset: Critical
exponent of the magnetization as a function of the defect intensity for $\epsilon=0$. The dashed line represents the analytic result (\ref{eq:bariev})}
\label{fig:mls_vs_L_e0}
\end{figure}

To begin with, we consider the decoupled model and compute the critical local magnetization of the $\sigma$ Ising plane having a ladder defect
\cite{Igloi_etal}. In this case $K_4=0$ implies that $K=K_c^{\textrm{Ising}}$, with $\sinh 2 K_c^{\textrm{Ising}}=0$. Fig. \ref{fig:mls_vs_L_e0} shows the
magnetization as a  function of $L$ (in logarithmic scale)  and we observe that the size dependence of the exponents is
negligible. The finite-size magnetization exponents are plotted in the inset as a function of defect intensity, where we observe a complete agreement with
the analytic result (\ref{eq:bariev}). The critical exponent for the $\tau$ spins is independent of $K_l$ and takes the value $x_{m}^{\tau}=\frac{1}{8}$
as expected.

\subsection{Positive coupling: $\epsilon>0$}

\begin{figure}[h!]
\begin{center}
\includegraphics[width=\figwidth]{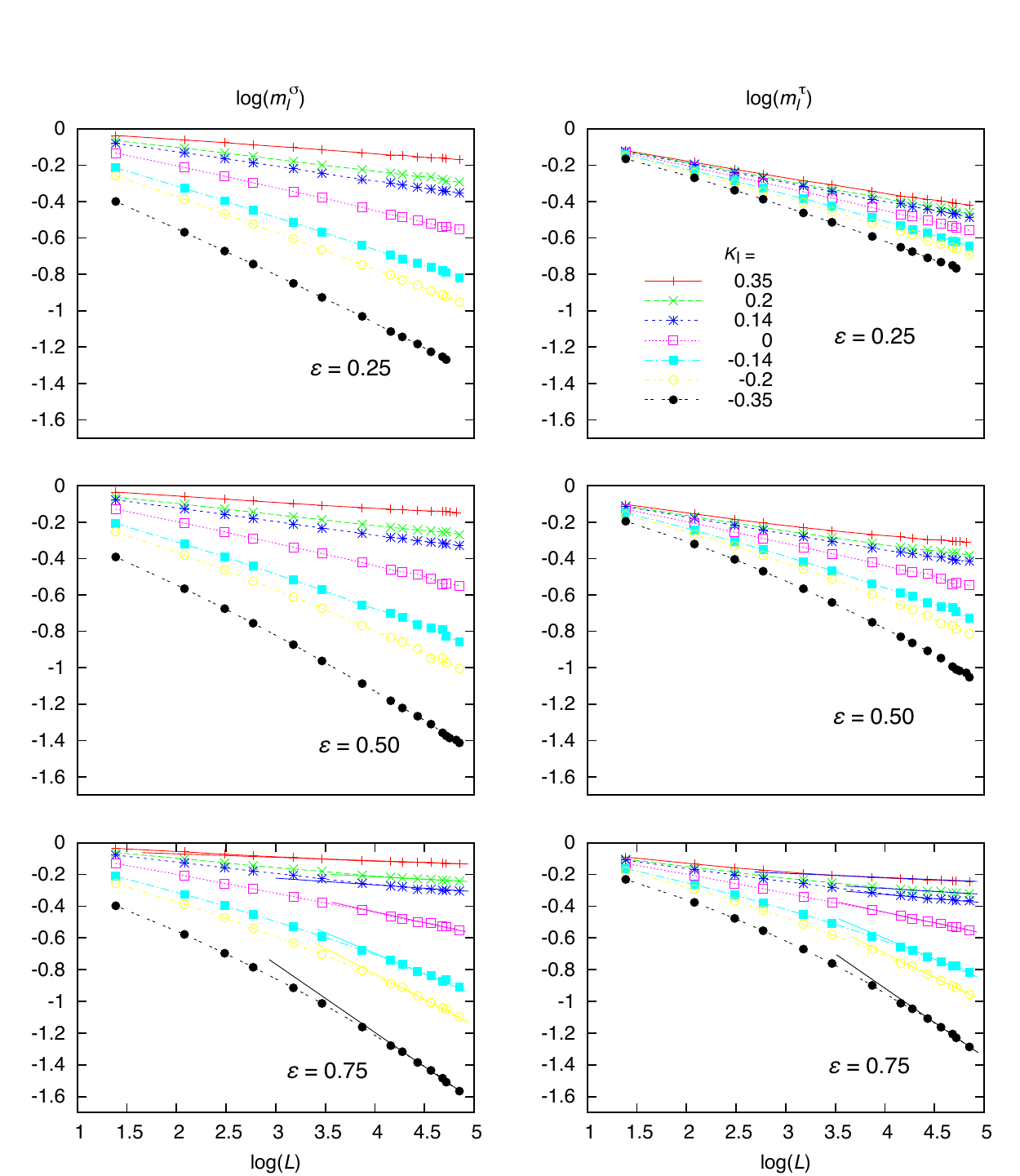}
\end{center}
\caption{Defect magnetization as a function of $L$ (in logarithmic scale) for several values of the coupling between Ising planes. In the left columns are
shown results for $\sigma$ spins whereas those for $\tau$ spins appear on the right. The slope is lower for positive values of $K_l$. }
\label{fig:ml_vs_L_all}
\end{figure}

When the coupling between $\sigma$ and $\tau$ spins is non-vanishing, the defect magnetization does not behave exactly as a power law, or in other words,
the exponents $x_{m}^{\sigma}$ y $x_{m}^{\tau}$ keep a residual $L$ dependence. This can be appreciated in Fig. \ref{fig:ml_vs_L_all} where we show the
defect magnetization of $\sigma$ and $\tau$ spins as a function of $L$ for different values of $K_l$ and $\epsilon$. One clearly observes deviations from
power laws. Still, for values of  $L$ larger than 64 (for $\epsilon=0.75$) a linear dependence in logarithmic scale is approached and we use fits in this
range to extract the exponents. For smaller values of $\epsilon$ the power law behaviour starts at smaller values of $L$. We observe that the slope at
large sizes increases with the value of $K_l$. The curve corresponding to the clean system $K_l=0$ has in all cases a slope close to $\frac{1}{8}$ and
shows very little deviations from that value.

\begin{figure}[h!]
\begin{center}
\includegraphics[width=\figwidth]{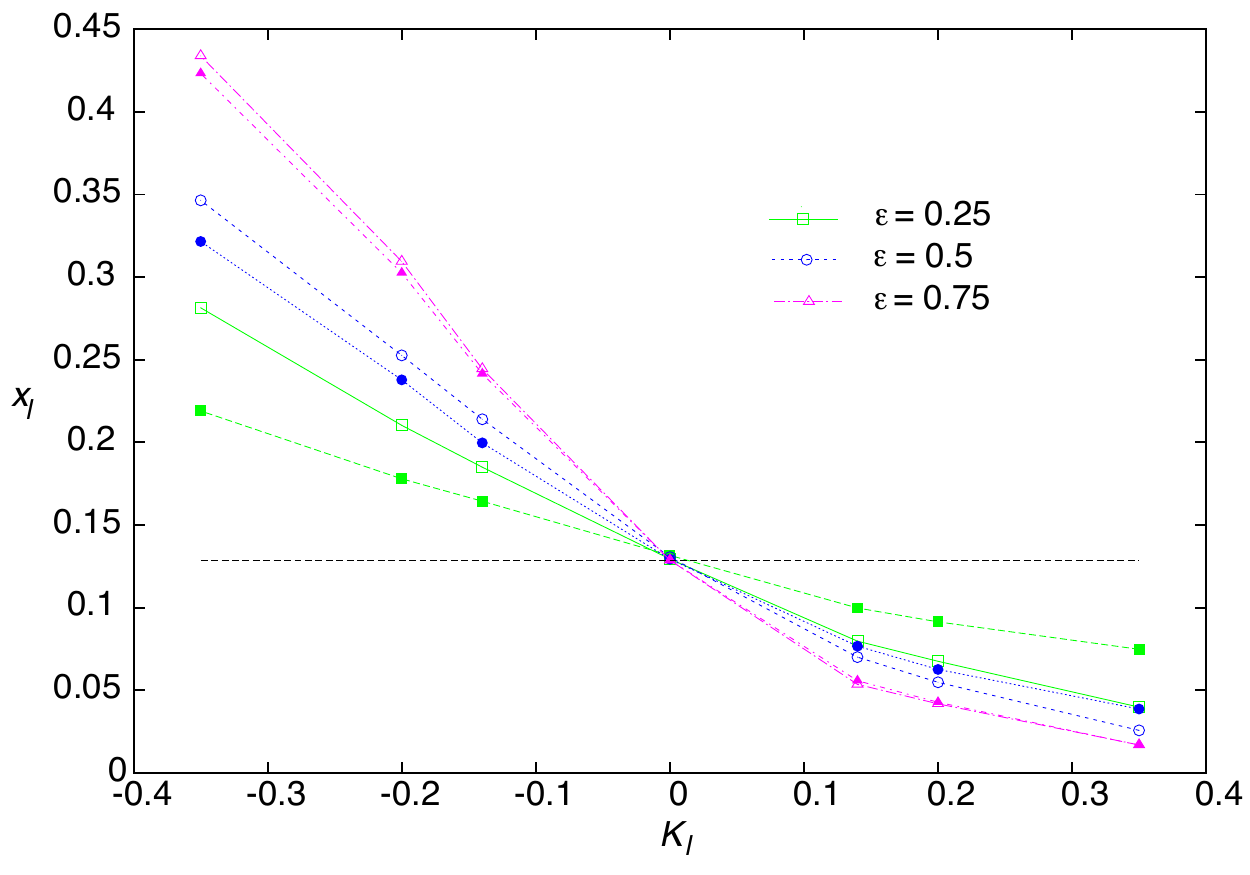}
\end{center}	
\caption{Critical exponent of the defect magnetization as a function of the defect intensity $K_l$. Results are shown for $\epsilon=0.25,\,0.5\,0.75$.
Filled points correspond to $x_l^\tau$ and empty points to $x_l^\sigma$. The horizontal slashed line signals the value of the exponents in absence of
defects, $x_l^\tau=x_l^\sigma=1/8$.}
\label{fig:exps_st_all}
\end{figure}

The behaviour of the critical exponents with the intensity of the defect is shown in Fig. \ref{fig:exps_st_all} for three values of the coupling
$\epsilon$. Deviations from the decoupled case increase with $\epsilon$ and are stronger for $x_l^\sigma$. Notice that for large values of $\epsilon$ the
difference between $x_l^\sigma$ and $x_l^\tau$ significantly reduces and seems to vanish for very large coupling. For positive $K_l$, exponents show lower
values than in the clean plane and tend to zero for large and positive defect intensities. This can be explained in terms of the phase
transition taking place in the system and its effects on the defect line. The spins lying on this line are coupled among them by an effective constant
$K_\mathrm{eff}=K+K_{l}$ that is stronger than the coupling with the spins in the bulk. Thus,  there is a tendency on the defect to order for values of
$K$ smaller than the critical value of the bulk $K_c$ and the defective line already finds itself in a sort of ``quasi ordered" state when the bulk still
transits from disorder to order. The local order is reflected in a smaller critical exponent. By the same reasoning, when the defect intensity is
negative, the effective value of the coupling among spins over the defective line is smaller than the coupling in the bulk and therefore the defective
line finds still disordered or in transition to order for $K$ equal to the critical value $K_c$ in the bulk.

\subsection{Negative coupling: $\epsilon<0$}

\begin{figure}[h!]
\begin{center}
\includegraphics[width=\figwidth]{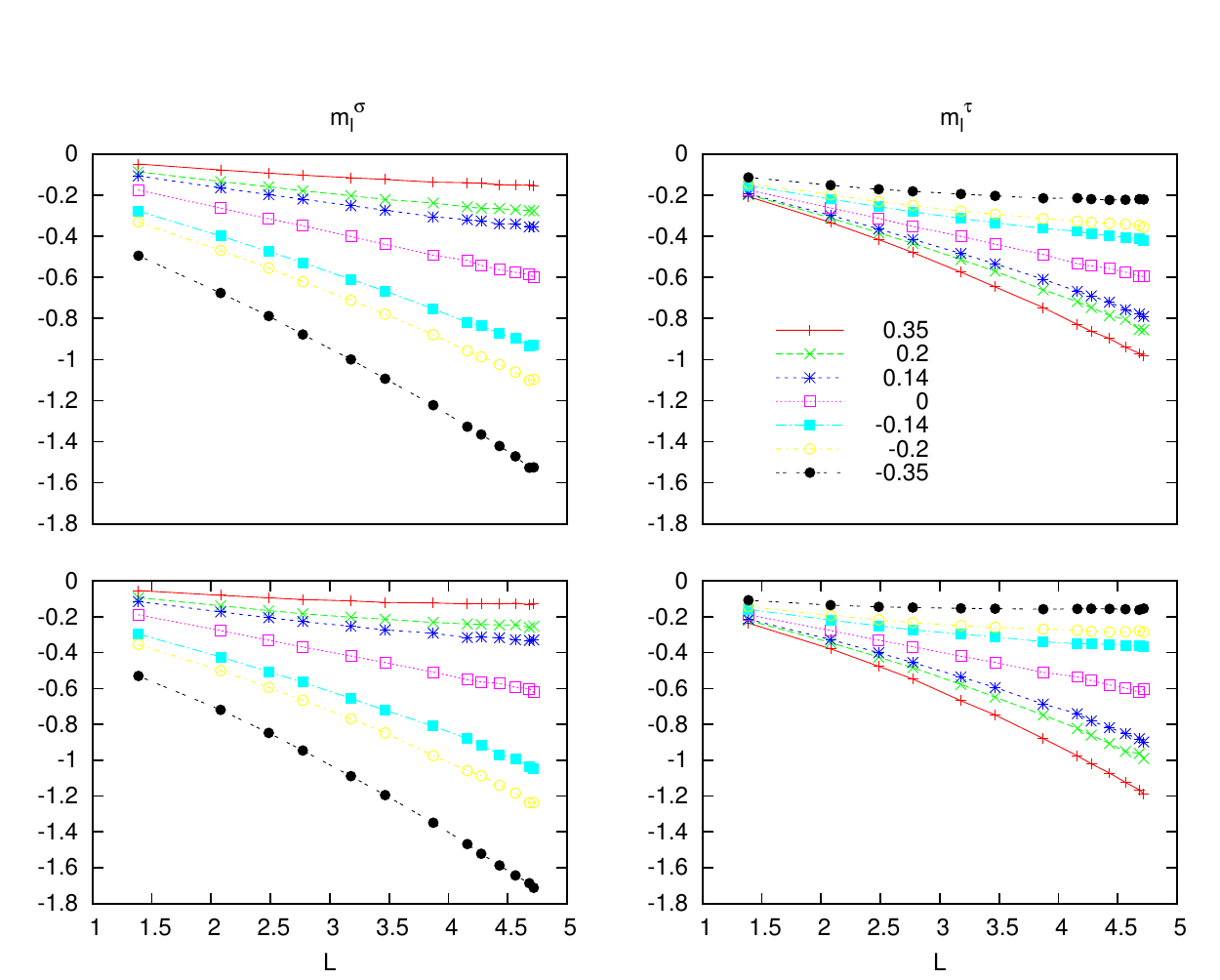}
\end{center}
\caption{Finite-size defect magnetization $m_{l}^{\alpha}$ for $\alpha=\sigma$ spins (left panel) and $\alpha=\tau$ spins (right panel).}
\label{fig:ml_vs_L_neg_all}
\end{figure}

\begin{figure}[h!]
\begin{center}
\includegraphics[width=\figwidth]{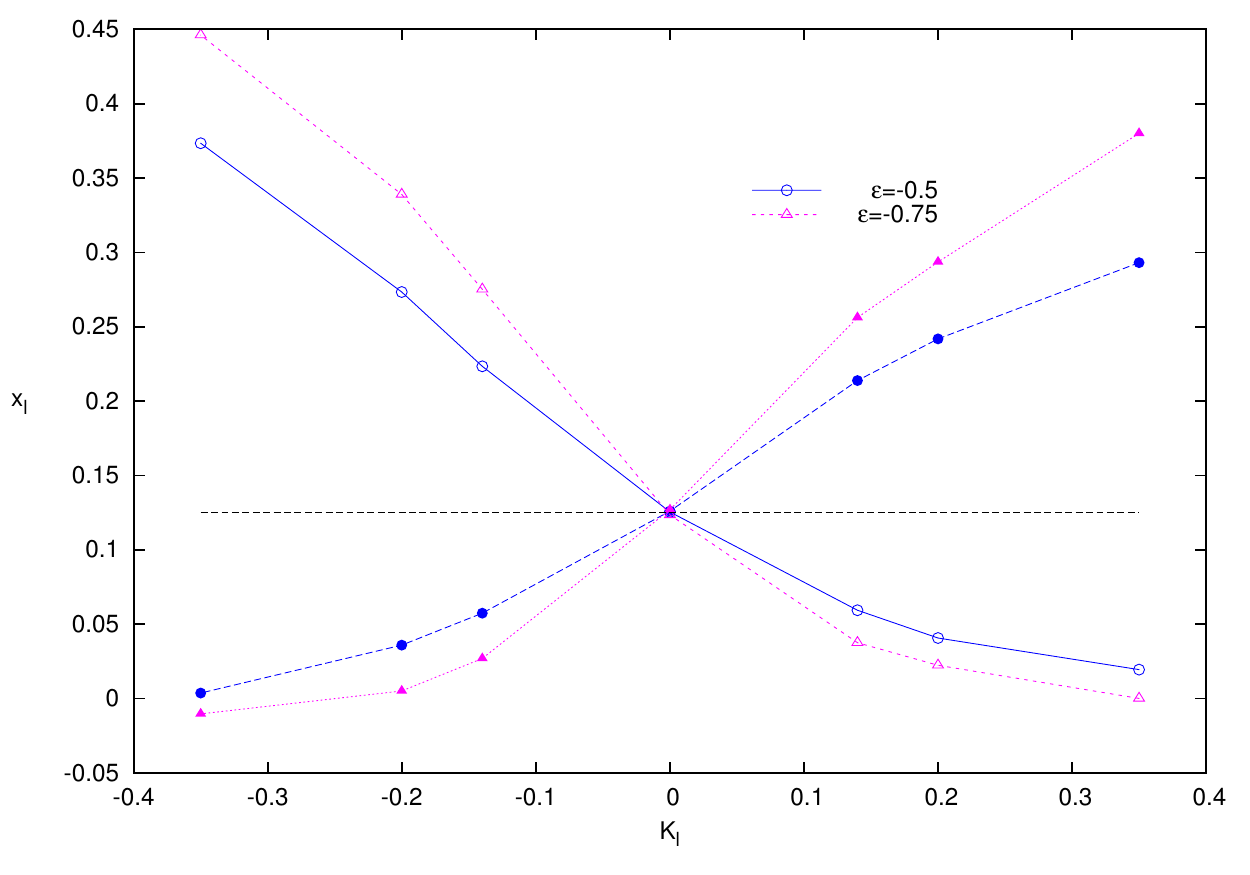}
\end{center}
\caption{Critical exponent of the defect magnetization as a function of the defect intensity for $\epsilon<0$. Results are shown for
$\epsilon=-0.5,\,-0.75$. Filled points correspond to $x_l^\tau$ and empty points to $x_l^\sigma$. The horizontal slashed line signals the value of the
exponents in absence of defects, $x_l^\tau=x_l^\sigma=1/8$. Notice the difference in the behaviour of the exponents associated to $\sigma$ and $\tau$
spins. The former decreases whereas the latter increases with the defect intensity.}
\label{fig:exp_neg_e0.75}
\end{figure}

The finite-size magnetization curves in this region of the phase diagram are exposed in Fig. \ref{fig:ml_vs_L_neg_all} and the calculated exponents are
shown in Fig.~\ref{fig:exp_neg_e0.75}. We observe in this case that the
behavior of the exponents for $\sigma$ and $\tau$ spins is different. The defect magnetization of the $\sigma$ spins follows the same behavior as in the
positive coupling case, and decreases with the intensity of the defect. On the other hand, the  magnetization of the $\tau$ spins is the reverse one, it
monotonically increases with the defect magnitude and tend to zero for $K_l$ large and negative.

\section{Conclusions}
In this paper we have reconsidered the computation of magnetic critical exponents in the Ashkin-Teller model with an asymmetric line defect. The main
motivation for this analysis is the discrepancy between the analytical results of Ref. \cite{AT_naon} and the numerical findings obtained in
\cite{AT_lajko}. Our results are in well agreement with this last work. In particular, for four-spin coupling $\epsilon >0$, which corresponds to the case studied in \cite{AT_naon}, we get $\sigma$-critical exponents that depend on both $\epsilon$ and $K_l$. Moreover, the form of this dependence is analogous to the one presented in \cite{AT_lajko} (See Fig.\ref{fig:exps_st_all}). On the other hand, the analytical calculation, based on functional integrals, gives critical exponents which are independent of $\epsilon$, i.e. only depends on $K_l$.  In view of these results it becomes natural to ask oneself about the reason for this disagreement. Let us recall that the method employed in \cite{AT_naon} is based on the evaluation of a fermionic determinant, which involves a regularization procedure. If one uses a gauge invariant prescription, as done in \cite{AT_naon}, then no $\epsilon$-dependent contribution to the determinant appears. And this is so because it is precisely the four-spin coupling in the Hamiltonian which breaks gauge invariance. This means that when performing the path-integral computation a general regularization containing $\epsilon$-dependent counterterms should be used, instead of the gauge invariant choice made in \cite{AT_naon}. Following this idea one could reconcile both numerical and analytical results for this problem. The details of this procedure will be worked out elsewhere.

\vspace{1cm}

\acknowledgments

This work was partially supported by CONICET (PIP 0662), ANPCyT (PICT
2010-1907)
and UNLP (PID X497), Argentina.


%

\end{document}